\documentclass[11pt,twoside]{article}

\usepackage{asp2014}

\aspSuppressVolSlug
\resetcounters

\begin{document}

\title{Classification of HI Galaxy Profiles Using Unsupervised Learning and Convolutional Neural Networks: A Comparative Analysis and Methodological Cases of Studies}

% Note the position of the comma between the author name and the 
% affiliation number.
% Authors surnames should come after first names or initials, eg John Smith, or J. Smith.
% Author names should be separated by commas.
% The final author should be preceded by "and".
% Affiliations should not be repeated across multiple \affil commands. If several
% authors share an affiliation this should be in a single \affil which can then
% be referenced for several author names. If only one affiliation, no footnotes are needed.
% See ManuscriptInstructions.pdf and ASP's manual2010.pdf 3.1.4 for more details
\author{Gabriel Jaimes-Illanes,$^1$ Manuel Parra-Royon,$^1$, Laura Darriba-Pol$^1$, Javier Moldón$^1$, Amidou Sorgho$^1$, Susana Sánchez-Expósito$^1$, Julián Garrido-Sánchez$^1$ and Lourdes Verdes-Montenegro$^1$}
\affil{$^1$Instituto de Astrofísica de Andalucía (IAA-CSIC), Granada, Spain; \email{gjaimes@iaa.es}
\email{mparra@iaa.es}
\email{ldarriba@iaa.es}
\email{jmoldon@iaa.es}
\email{asorgho@iaa.es}
\email{sse@iaa.es}
\email{jgarrido@iaa.csic.es}
\email{lourdes@iaa.es}}

% This section is for ADS Processing.  There must be one line per author. paperauthor has 9 arguments.
\paperauthor{Gabriel Jaimes}{gjaimes@iaa.es}{0000-0002-3339-4128}{Instituto de Astrofísica de Andalucía (IAA)}{{Extragalactic Department - AMIGA Group}}{Granada}{18008}{Spain}

\paperauthor{Manuel Parra}{mparra@iaa.es}{0000-0002-6275-8242}{Instituto de Astrofísica de Andalucía (IAA)}{Extragalactic Department - AMIGA Group}{Granada}{Andalucia}{18008}{Spain}
\paperauthor{Laura Darriba}{ldarriba@iaa.es}{0000-0002-5599-2647}{Instituto de Astrofísica de Andalucía (IAA)}{Extragalactic Department - AMIGA Group}{Granada}{Andalucia}{18008}{Spain}
\paperauthor{Laura Darriba}{ldarriba@iaa.es}{0000-0002-5599-2647}{Instituto de Astrofísica de Andalucía (IAA)}{Extragalactic Department - AMIGA Group}{Granada}{Andalucia}{18008}{Spain}
\paperauthor{Javier Moldón}{jmoldon@iaa.es}{0000-0002-8079-7608}{Instituto de Astrofísica de Andalucía (IAA)}{Extragalactic Department - AMIGA Group}{Granada}{Andalucia}{18008}{Spain}
\paperauthor{Amodou Shorgo}{asorgho@iaa.es}{0000-0002-5233-8260}{Instituto de Astrofísica de Andalucía (IAA)}{Extragalactic Department - AMIGA Group}{Granada}{Andalucia}{18008}{Spain}
\paperauthor{Susana Expósito}{sse@iaa.es}{0000-0002-7510-7633}{Instituto de Astrofísica de Andalucía (IAA)}{Extragalactic Department - AMIGA Group}{Granada}{Andalucia}{18008}{Spain}
\paperauthor{Julián Garrido}{jgarrido@iaa.csic.es}{0000-0002-6696-4772}{Instituto de Astrofísica de Andalucía (IAA)}{Extragalactic Department - AMIGA Group}{Granada}{Andalucia}{18008}{Spain}
\paperauthor{Lourdes Verdes-Montenegro}{lourdes@iaa.es}{000-0003-0156-6180}{Instituto de Astrofísica de Andalucía (IAA)}{Extragalactic Department - AMIGA Group}{Granada}{Andalucia}{18008}{Spain}

% There should be one \aindex line (commented out) for each author. These are used to
% build up the author index for the Proceedings. The surname must come first, followed by
% initials. Note the use of ~ before each initial to control spacing.
% The \author entries (see above) have surname last. These \aindex entries have
% surname first.
% The Aindex.py command willl create them for you after you have constructed the \author
% The first entry should be the first author, for bold-facing the author index page-reference

%\aindex{FistAuthor1,~S.~A.}
%\aindex{Author2,~S.~B.}
%\aindex{Author3,~S.}

\begin{abstract}

Hydrogen, the most abundant element in the universe, is crucial for understanding galaxy formation and evolution. The 21 cm neutral atomic hydrogen (HI) spectral line maps the gas kinematics within galaxies, providing key insights into interactions, galactic structure, and star formation processes. With new radio instruments, the volume and complexity of data is increasing. To analyze and classify integrated HI spectral profiles in a efficient way, this work presents a framework that integrates Machine Learning (ML) techniques, combining unsupervised methods and Convolutional Neural Networks (CNN). To this end, we apply our framework to a selected subsample of 318 spectral HI profiles of the catalogue of Isolated Galaxies (CIG) and 30.780 profiles from the Arecibo Legacy Fast ALFA Survey (\mbox{ALFALFA}) catalogue. Data pre-processing involved the \texttt{Busyfit} package and iterative fitting with polynomial, Gaussian, and double-Lorentzian models. Clustering methods, including K-means, spectral clustering, \mbox{DBSCAN}, and agglomerative clustering, were used for feature extraction and to bootstrap classification we applied K-Nearest Neighbors, Support Vector Machines, and Random Forest classifiers, optimizing accuracy with CNN. Additionally, we introduced a 2D model of the profiles to enhance classification by adding dimensionality to the data. Three 2D models were generated based on transformations such as rotation, subtraction of profiles, and normalised versions to quantify the level of asymmetry. These methods were tested in a previous analytical classification study conducted by the Analysis of the Interstellar Medium in Isolated Galaxies group (AMIGA). This approach enhances classification accuracy and aims to establish a methodology that could be applied to data analysis in future surveys conducted with the Square Kilometre Array (SKA), currently under construction. All materials, code, and models have been made publicly available in an open-access repository, adhering to FAIR  principles.

\end{abstract}

% These lines show examples of subject index entries. At this stage these have to commented
% out, and need to be on separate lines. Eventually, they will be automatically uncommented
% and used to generate entries in the Subject Index at the end of the Proceedings volume.
% Don't leave these in! - replace them with ones relevant to your paper.
%\ssindex{FOOBAR!conference!ADASS 2019}
%\ssindex{FOOBAR!organisations!ASP}

% These lines show examples of ASCL index entries. At this stage these have to commented
% out, and need to be on separate lines. Eventually, they will be automatically uncommented
% and used to generate entries in the ASCL Index at the end of the Proceedings volume.
% The ascl.py command will scan your paper on possible code names.
% Don't leave these in! - replace them with ones relevant to your paper.
%\ooindex{FOOBAR, ascl:1101.010}

\section{Introduction}
%\subsection{Neutral atomic hydrogen (HI)}
% Lo cambio completamente por tener muchas repeticiones de algunos elementos:
%The study of the interstellar medium (ISM), structures and dynamics, reveals complex phenomena that contribute to the comprehension of galaxies evolution. Hydrogen is the most abundant element in the universe, playing a critical role in the formation and evolution of galaxies. The 21 cm radio wavelength neutral atomic hydrogen (HI) line maps the distribution and dynamics of gas within galaxies. The emission from this spectral line is an important tracer for galaxy interaction studies and understanding galactic structure, star formation processes and general behavior of the Interstellar Medium. This emission serves as a tracer for interactions between galaxies, in addition to the study of galactic structure, star formation processes and the overall behavior of the interstellar medium.
The study of the interstellar medium (ISM) within galaxies, and the intergalactic medium (IGM) surrounding them, reveals complex phenomena that are fundamental for understanding galaxy evolution. Hydrogen, the most abundant element in the universe, plays a pivotal role in these processes. The neutral atomic hydrogen (HI) line, observed at a 21-cm radio wavelength, provides critical insights into the distribution and dynamics of gas within galaxies. This spectral line is a key tracer for galaxy interactions, galactic structure, star formation processes, and the overall behavior of the ISM. This information provided by HI observations could be addressed through ML techniques to improve the efficiency of processing such data and allow more detailed studies of galactic dynamics and evolution.%\subsection{Profiles classification using machine learning}
The application of ML, BigData tools, and algorithms are assets to tackle the enhancement of the quality in scientific analysis. %especially when it involves large radio astronomy databases and the study of a classification problem with HI spectrum. 
Within the AMIGA research group (\url{https://amiga.iaa.csic.es/}), our work aims to propose a framework to classify HI spectral profiles through ML techniques using HI datasets. 
%In this context, our work aims to propose a framework for the classification of HI spectral profiles using ML techniques. Several methodologies have been implemented including unsupervised ML techniques and Convolutional Neural Networks (CNN). To carry out this approach, we worked on the HI datasets used in the AMIGA research group, covering 318 profiles from the CIG sample\footnote{catalogue of Isolated Galaxies (CIG) (Karachentseva et al., 1986) \url{https://vizier.cds.unistra.fr/viz-bin/VizieR?-meta.foot&-source=VII/82A}.}  and 30.780 spectrum profiles from the ALFALFA survey\footnote{The Arecibo Legacy Fast ALFA Survey (ALFALFA) of HI extragalactic sources (Haynes et al., 2018) \url{https://egg.astro.cornell.edu/alfalfa/data/}.}.

This work is conducted within the context of the Spanish prototype of a SKA Regional Centre espSRC \citep[][]{2023hsa..conf..360G}.

\section{Methodology}
\subsection{Spectral line data and Pre-Processing}
%The classification framework, as shown in Figure \ref{ex_fig1}, considered as first step the data preprocessing. First, a second data set was generated using iterative fitting with polynomial, Gaussian, and double-Lorentzian models. Secondly, another dataset using the Busyfit package for "Busy Function"  \citep[][]{2014MNRAS.438.1176W} for HI spectrum profile fitting. 

The spectral line data include 318 profiles from the CIG sample \cite[][]{1986BICDS..30..125K}\footnote{Catalogue of Isolated Galaxies (CIG) \url{https://vizier.cds.unistra.fr/viz-bin/VizieR?-meta.foot&-source=VII/82A}.}, showing intensity versus velocity, used for asymmetry analysis and classification, limited in quantity due to comparison with previous analytical methods. Secondly, 30.780 spectrum profiles from the ALFALFA survey \cite[][]{2018ApJ...861...49H}\footnote{The Arecibo Legacy Fast ALFA Survey (ALFALFA) of HI extragalactic sources \url{https://egg.astro.cornell.edu/alfalfa/data/}.}, presenting flux versus heliocentric velocity and parameters such as HI line flux density, peak flux-to-noise ratio, adopted distance, and uncertainties, were used for multiple classifications with ML techniques. Figure \ref{ex_fig1} illustrates the classification methodology, starting with data pre-processing. Useful datasets were generated through iterative fitting with polynomial, Gaussian and double lorentz models, and using the \texttt{Busyfit} package with the "Busy function" \citep[][]{2014MNRAS.438.1176W} for the fitting of the HI spectrum.

\subsection{Classification methods analysis}
%The first part of the classification was focused on ALFALFA catalogue, which involved a multi-faceted strategy for profile clustering based on temporal shapelets transformation for features detection algorithms: K-means, spectral clustering, DBSCAN, agglomerative clustering, among others, as bootstrap for the extraction of features.
%Following, we considered a series of classification techniques that include K-Nearest Neighbors (KNN), Support Vector Machines (SVM), and Random Forest classifiers. In order to optimize the performance of such models, CNN model was probed, where we made an in depth evaluation for various configurations of the model with regard to their impact on classification accuracy.

The classification process began with the ALFALFA catalogue, using clustering for pattern identification and bootstrapping for feature extraction. Clustering methods such as K-means, spectral clustering, DBSCAN, and agglomerative clustering were applied using the temporal shapelet transform. Subsequently, classification techniques such as K-Nearest Neighbors (KNN), Support Vector Machines (SVM), and Random Forest were explored. To optimize performance, a CNN model was tested with various configurations to evaluate its impact on accuracy.

\articlefigure{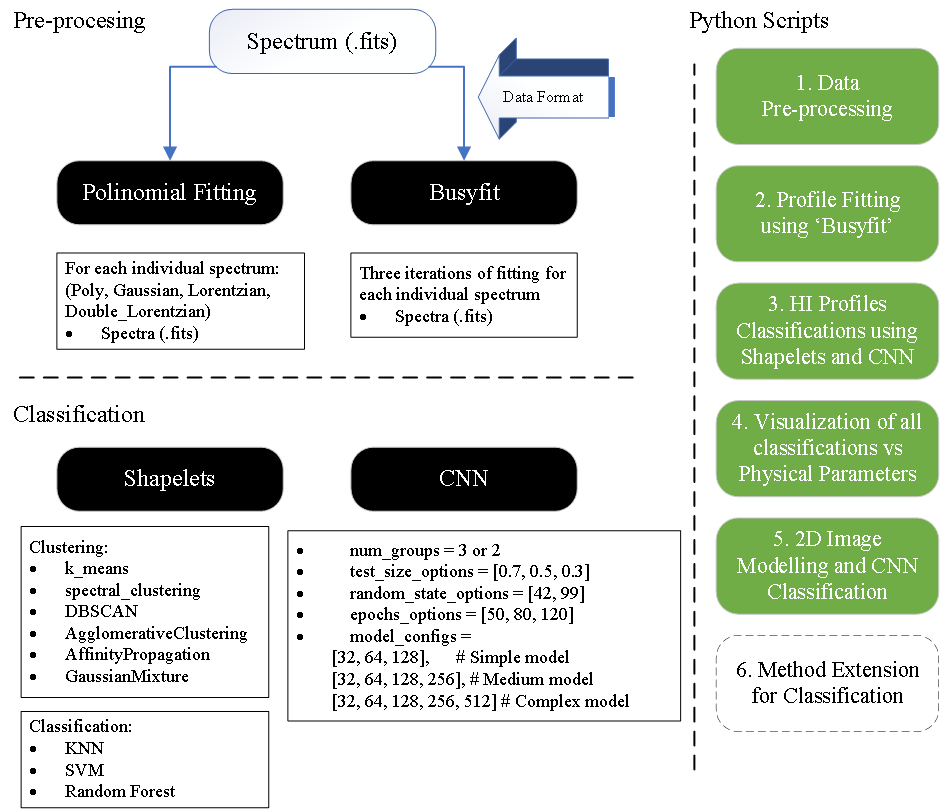}{ex_fig1}{Diagram for the selected HI profile classification methodology including pre-processing, classification, and developed code with \texttt{Python}.}

\subsection{Asymmetry analysis using 2D modeling}

%The second part of this work is focused on CIG AMIGA sample of galaxies. Therefore, the generation of an additional dimension to the profiles in order to improve the classification. This 2D analysis was based on the application of CNN techniques to determine the degree of asymmetry by carrying out the classification of the sample of CIG galaxies. The original data was modified by adding a new dimension to the profiles in order to improve the classification. Three distinct 2D image models were generated for the symmetry study: 

Another part of the work focused on the AMIGA sample of CIG galaxies, with the aim of improving the classification by adding an extra dimension to the HI profiles. This 2D analysis and modeling used CNN techniques to assess the degree of asymmetry and classify the sample of CIG galaxies. The original data were transformed by introducing a new dimension to the profiles, which allowed us to improve the classification. 

For the asymmetry study, three 2D image models were generated: a) \texttt{Model 1}: Rotated spectrum, where the original profile was rotated to create a 2D representation; b) \texttt{Model 2}: Subtracted asymmetry model, reflecting one side of the split profile at its center, calculating the difference, and rotating it to highlight asymmetry; and c) \texttt{Model 3}: Normalized asymmetry model, a normalized version of \texttt{Model 2} with pixel intensity adjusted to enhance specific features.

The resulting classification was compared\footnote{Results: \url{https://github.com/gabojaimesillanes/JAE-Intro-ICU-2024-Gabriel-Jaimes-IAA/blob/95530cc8cd60e6d9f033dd1ebfcf71b526baee65/E.\%20Asymetry_Analysis_CIG_AMIGA_IAA.ipynb}} with an analytical profile classification \citep[][]{2011A&A...532A.117E} previously made by the AMIGA group.

\section{Results and products}

%This work is fully available in a public GitHub repository: \url{https://github.com/gabojaimesillanes/JAE-Intro-ICU-2024-Gabriel-Jaimes-IAA}. This includes documentation that provides a detailed guide, for experiments and reproduction of results by the usage of the classification methodology applied, which can be accessed at: \url{https://astrodev-g.readthedocs.io/en/latest/1.HI_class/}.

%A methodological framework has been implemented, incorporating unsupervised techniques and CNN to analyze the profiles from the CIG and ALFALFA catalogues (see Figure \ref{ex_fig2}). This process involved data pre-processing using the BusyFit package, iterative fittings with polynomial, Gaussian, and Lorentzian models, and the application of classifiers such as KNN, SVM, and Random Forest.

%A methodological framework was developed and successfully implemented to analyze the profiles from the CIG and ALFALFA catalogues (see Figure \ref{ex_fig2}). The results of this approach incorporate unsupervised techniques and CNNs to process the data. 

A methodological framework incorporating unsupervised techniques and CNNs to process the data was developed and successfully implemented to analyze ALFALFA catalogue profiles (see Figure \ref{ex_fig2}).
The analysis involved pre-processing steps using the \texttt{Busyfit} package, followed by iterative fittings with polynomial, Gaussian, and Lorentzian models. Subsequently, classifiers such as KNN, SVM, and Random Forest were applied. The results give a better understanding of the classification of galaxy profiles and demonstrate the effectiveness of this methodology.

\articlefiguretwo{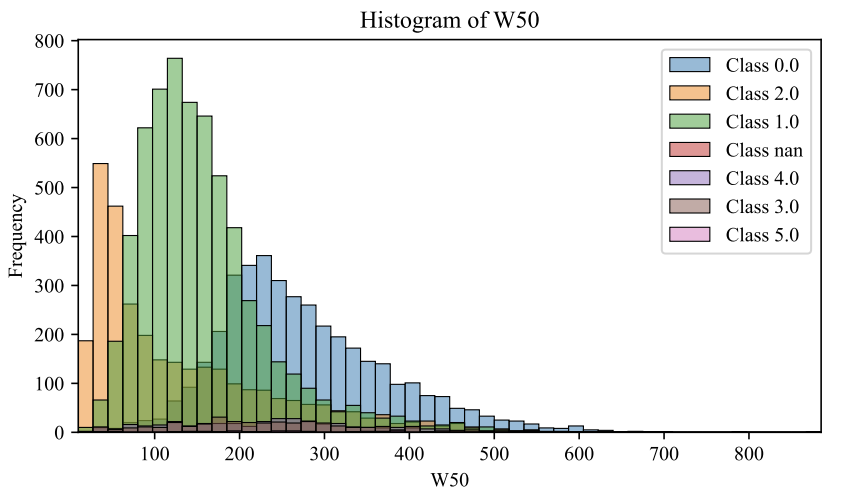}{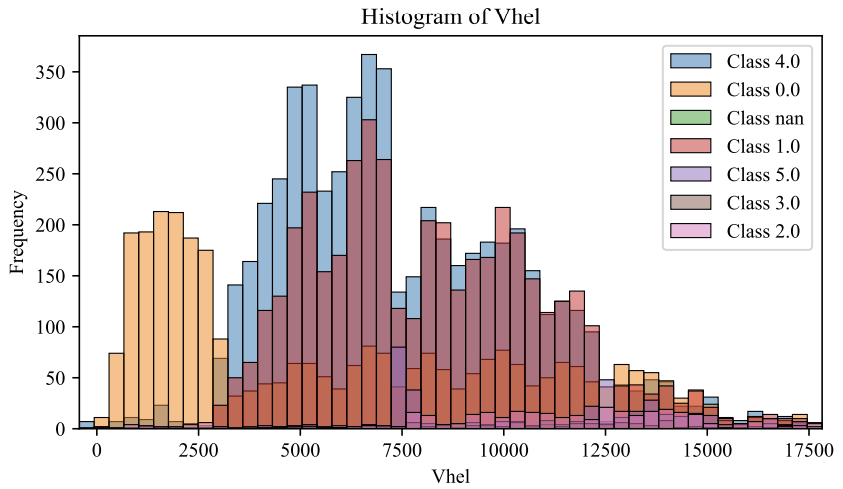}{ex_fig2}{Histograms of classifications as a function of Width at 50 percent (W50) and Heliocentric velocity (VHel), presented in the left and right panels, respectively.}

We obtained a classification success rate of 63\% for classifying profiles with 2D images. The study highlighted the adaptability of the methodology to various configurations and parameters. Different test sizes, random states, and training epochs were explored. Furthermore, 54 classification configurations were performed for each iteration, allowing thorough debugging and comparison with the previous AMIGA group work. A key challenge encountered was the computational resources required, with an average time of 1.13 hours per iteration on a system with 16 GB of RAM and 8 CPUs. Despite this, the results suggest that the methodology is robust and scalable.

\articlefigure[width=.8\textwidth]{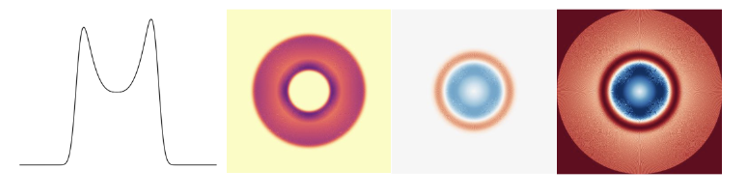}{ex_fig5}{Emission line spectrum shape, Model 1, Model 2, and Model 3, shown from left to right.}

%The results include a classification success rate of 63\% for classifying profiles with 2D images. 

This work is fully accessible through a public \texttt{GitHub} repository\footnote{Repository: \url{https://github.com/gabojaimesillanes/JAE-Intro-ICU-2024-Gabriel-Jaimes-IAA}}. This repository includes complete documentation with detailed guidance for conducting experiments and reproducing results using the applied classification methodology. The documentation can be accessed at \url{https://astrodev-g.readthedocs.io/en/latest/1.HI_class/}.

\

\section{Conclusions and future work}

This study highlights the potential of combining ML techniques, including CNNs and Shapelet transformations, with radio astronomy data to classify HI spectral profiles effectively. The ALFALFA data classifications, visualized against physical parameters, provide additional information on the performance of the 18 classifications conducted. Transforming 1D sample profiles into 2D images has proven to be a robust method, improving asymmetry detection and improving classification accuracy by 13\% compared to traditional 1D methods. These results demonstrate the potential of advanced computational tools in extracting valuable information from large-scale astronomical datasets, paving the way for comprehensive studies of galaxy dynamics and evolution. Future work will aim to refine 2D transformations, optimize CNN architectures, and expand the framework to incorporate new datasets. Within the context of the project in which this work is framed, the application of this approach in SKA-era surveys could allow the efficient classification of millions of HI profiles. 

%This project establishes a solid foundation for addressing the challenges of next-generation astronomical observations. 

\acknowledgements This work was carried out under the JAE Intro ICU 2023 Research Scholarship in the AMIGA group within the Instituto de Astrofísica de Andalucía, Consejo Superior de Investigaciones Científicas (CSIC). Authors acknowledge financial support from the grant CEX2021-001131-S funded by MICIU/AEI/ 10.13039/ 501100011033. MP, LD, JM, AS, SSE, JGS and LVM acknowledge financial support from the grant PID2021-123930OB-C21 funded by MICIU/AEI/ 10.13039/501100011033 and by ERDF/EU and from the grant TED2021-130231B-I00 funded by MICIU/AEI/ 10.13039/501100011033 and by the European Union NextGenerationEU/PRTR. JM acknowledges financial support from grant PID2023-147883NB-C21, funded by MCIU/ AEI/  10.13039/501100011033 and by ERDF/EU. The authors acknowledge the Spanish Prototype of an SRC (SPSRC) service and support funded by the Ministerio de Ciencia, Innovación y Universidades (MICIU), by the Junta de Andalucía, by the European Regional Development Funds (ERDF) and by the European Union NextGenerationEU/PRTR. The SPSRC acknowledges financial support from the Agencia Estatal de Investigación (AEI) through the "Center of Excellence Severo Ochoa" award to the Instituto de Astrofísica de Andalucía (IAA-CSIC) (SEV-2017-0709) and from the grant CEX2021-001131-S funded by MICIU/AEI/ 10.13039/501100011033.
%This work was carried out under the JAE Intro ICU 2023 Research Scholarship in the AMIGA group within the Instituto de Astrofísica de Andalucíaa (IAA), Consejo Superior de Investigaciones Científicas (CSIC). Also, with the support of grants PID2021-123930OB-C21, TED4SKA, PRE2021-100660, CEX2021-001131-S and INFRA24023 (CSIC4SKA). MP, LD, SS, JG and LVM acknowledge financial support from the grant PID2021-123930OB-C21 funded by MICIU/AEI/ 10.13039/ 5011 00011033 and by ERDF/EU. MP, LD, SS, JG and LVM acknowledge financial support from the grant TED2021-130231B-I00 funded by MICIU/AEI/ 10.13039/501100011033 and by the European Union NextGenerationEU/PRTR. JM acknowledges financial support from grant PID2023-147883NB-C21, funded by MCIU/AEI/ 10.13039/501100011033 and by ERDF/EU. The authors acknowledge the Spanish Prototype of an SRC (ESPSRC) service and support funded by the Ministerio de Ciencia, Innovación y Universidades (MICIU), by the Junta de Andalucía, by the European Regional Development Funds (ERDF) and by the European Union NextGenerationEU/PRTR. The ESPSRC acknowledges financial support from the Agencia Estatal de Investigación (AEI) through the "Center of Excellence Severo Ochoa" award to the IAA-CSIC (SEV-2017-0709) and from the grant CEX2021-001131-S funded by MICIU/AEI/ 10.13039/501100011033.

\bibliography{bi1,bi2,bi3,bi4,b5}

\end{document}